\documentclass[journal,twoside,web]{ieeecolor}
\usepackage{generic}
\usepackage{cite}
\usepackage{amsmath,amssymb,amsfonts}
\usepackage{algorithmic}
\usepackage{graphicx}
\usepackage{textcomp}
\usepackage{color}
\usepackage{soul}
\usepackage{textcomp}
\usepackage{comment}
\usepackage{xcolor}
\def\BibTeX{{\rm B\kern-.05em{\sc i\kern-.025em b}\kern-.08em
 T\kern-.1667em\lower.7ex\hbox{E}\kern-.125emX}}

\begin{document}
\title{HDE-Array: Development and Validation of a New Dry Electrode Array Design to Acquire HD-sEMG for Hand Position Estimation}\author{Giovanni Rolandino, \IEEEmembership{Student Member, IEEE}, Chiara Zangrandi, Taian Vieira,\\ Giacinto Luigi Cerone, \IEEEmembership{Member, IEEE} , Brian Andrews and James J. FitzGerald, \IEEEmembership{Member, IEEE} \thanks{G. Rolandino, J. J. FitzGerald and B. Andrews are with the Nuffield Department of Surgical Sciences, University of Oxford, Oxford, OX3 9DU, UK (e-mail: giovanni.rolandino@nds.ox.ac.uk). C. Zangrandi, T. Vieira and G. L. Cerone are with LISiN (Department of Electronics and Telecommunications), Politecnico di Torino, Turin, 10129, Italy (e-mail: taian.martins@polito.it). This article has supplementary downloadable material available at https://zenodo.org/records/11217287, provided by the authors. Copyright (c) 2024 IEEE. Personal use of this material is permitted. However, permission to use this material for any other purposes must be obtained from the IEEE by sending an email to pubs-permissions@ieee.org.}}

\maketitle

\begin{abstract}
This paper aims to introduce HDE-Array (High-Density Electrode Array), a novel dry electrode array for acquiring High-Density surface electromyography (HD-sEMG) for hand position estimation through RPC-Net (Recursive Prosthetic Control Network), a neural network defined in a previous study. We aim to demonstrate the hypothesis that the position estimates returned by RPC-Net using HD-sEMG signals acquired with HDE-Array are as accurate as those obtained from signals acquired with gel electrodes. We compared the results, in terms of precision of hand position estimation by RPC-Net, using signals acquired by traditional gel electrodes and by HDE-Array. As additional validation, we performed a variance analysis to confirm that the presence of only two rows of electrodes does not result in an excessive loss of information, and we characterized the electrode-skin impedance to assess the effects of the voltage divider effect and power line interference. Performance tests indicated that RPC-Net, used with HDE-Array, achieved comparable or superior results to those observed when used with the gel electrode setup. The dry electrodes demonstrated effective performance even with a simplified setup, highlighting potential cost and usability benefits. These results suggest improvements in the accessibility and user-friendliness of upper-limb rehabilitation devices and underscore the potential of HDE-Array and RPC-Net to revolutionize control for medical and non-medical applications.
\end{abstract}


\section{Introduction}
\label{sec:Intro}
\IEEEPARstart{A}{s} of 2020, the annual incidence of upper limb amputations in High-Income regions was 13.5 per 100,000 people \cite{mcdonald2021global}\cite{lopez1998global}\cite{yuan2023global}. Spinal cord lesions at neck level, causing upper limb impairment, had an incidence of 10.7 per 100,000 in the same region \cite{Ding2022}\cite{NSCISC2022}\cite{singh2014global}. These figures underscore the urgent need for effective upper-limb rehabilitation technologies \cite{mohammed2014quality}. Significant advances have been made in this field, particularly in prostheses mimicking hand movements and Functional Electrical Stimulation (FES) to restore motion \cite{catalano2014adaptive}\cite{abryandoko2024literature}\cite{controzzi2016sssa}\cite{laffranchi2020hannes}\cite{fajardo2020galileo}. For those, various control mechanisms have been explored, with surface electromyography (EMG) showing the most promise \cite{Igual2019}\cite{Marinelli2023}\cite{Jiang2023}\cite{Mendez2021}. EMG's potential as a control source extends beyond rehabilitation to broader Human-Computer Interaction (HCI) applications, such as handwriting recognition for dyslexia treatment and injury rehabilitation \cite{tigrini2024intelligent}, as well as Augmented Reality (AR), Virtual Reality (VR), and secure identification using bio-markers \cite{kwon2021electromyography}\cite{sugiarto2021surface}\cite{jiang2022measuring}. EMG is also gaining interest in gaming and music \cite{hughes2021applications}\cite{fen2021virtual}. However, translating these applications to rehabilitative devices remains challenging due to patient-specific factors. For instance, amputees and tetraplegic patients may lack control over forearm muscles, which are often used in VR and gaming for consistent EMG signals \cite{botros2022day}. Still, cross-disciplinary research offers valuable insights, especially in enhancing post-processing algorithms and related techniques, and ultimately benefits all fields. Despite significant technological advances in the field of rehabilitation devices, abandonment rates, a key indicator of user satisfaction, remain high, often exceeding 30 \% \cite{salminger2022current}\cite{biddiss2007up}\cite{Jiang2023}\cite{Mendez2021}\cite{cordella2016literature}. Contributing factors include lack of comfort, functionality, and aesthetic appeal, with no single issue being dominant \cite{salminger2022current}\cite{smail2021comfort}\cite{kyberd2017survey}. Users frequently blame discomfort or pain, and the fact that control systems do not feel natural, a quality typically defined by independent finger movement, force control, ease of daily tasks, and sensory feedback integration \cite{biddiss2007upper}\cite{Kerner2021}\cite{ostlie2012prosthesis}\cite{Einfeldt2023}.

In a previous work, we addressed these issues with Recursive Prosthetic Control Network (RPC-Net), which estimates hand position using High-Density surface Electromyography (HD-sEMG) from the forearm \cite{artgio}. RPC-Net surpasses the performance of existing state-of-the-art control solutions and effectively addresses some of the issues that have historically hindered user acceptance and comfort, mostly related to naturalness of control. However, during the validation process we could appreciate how traditional gel electrodes, though effective at capturing EMG signals, have significant drawbacks. They require adhesive pads and conductive paste, making them cumbersome and not practical for daily use and complicating the acquisition setup. Building on RPC-Net's success, we developed a new solution to address these issues: the High-density Dry Electrode Array (HDE-Array). HDE-Array consists of two circumferences of 32 dry electrodes each, eliminating the need for adhesive pads and conductive paste required by gel arrays. This makes it more cost-effective, comfortable, and faster for users, improving accessibility for daily use and addressing a key factor in device acceptance. The growing preference for dry electrodes is reflected in recent literature \cite{li2017towards}. Additionally, the HDE-Array includes more electrodes per circumference than any current dry-electrode solution \cite{Marinelli2023} \cite{simpetru2023proportional}, providing unprecedented resolution crucial for capturing detailed EMG signals and aligning with the forearm's anatomical structure \cite{drake2012gray}. This higher spatial resolution aims to improve signal quality, enhancing control accuracy. However, while the RPC-Net/HDE-Array system is expected to improve comfort and usability of a rehabilitation device, the new hardware may introduce drawbacks, such as lower proximo-distal coverage and higher skin-electrode impedance compared to gel electrodes, which must be evaluated, especially in the way they affect the performance of RPC-Net.

The objective of the study presented in this paper is to assess how these potential drawbacks affect the performance of RPC-Net when using HD-sEMG signals acquired with HDE-Array, rather than with traditional gel electrode arrays. To do this, we test the hypothesis that the accuracy of hand position estimates in these two cases is equal. The goal is to show that RPC-Net can operate effectively with the HDE-Array, expanding its usability in rehabilitation solutions. Additionally, we characterized the skin-electrode impedance and compared it to the EMG amplifier input impedance and to the skin-electrode impedance for gel electrodes, analyzed the variance of the signal across rows and columns (to confirm that the proximo-distal coverage is sufficient), and evaluated the performance of RPC-Net using a single row of electrodes. To conduct these experiments, we acquired two datasets (each including HD-sEMG and hand position): one using dry electrodes and one using gel electrodes. The dataset with gel electrodes was used to assess changes in proximo-distal variance, while both datasets were used together to test overall performance. 

\section{Materials and Methods}
This section describes the two datasets used for the experiments (Section \ref{sec:DS}). We then detail the experimental approach employed in the three parts making up this study (Sections \ref{sec:variance} \ref{sec:impedance}, and \ref{sec:RPC}).

\subsection{Data}\label{sec:DS}
For this study, we used two HD-sEMG and hand kinematics recordings datasets. The EMG signal was used for the training and testing phase of RPC-Net while hand position data were used both as input and as target values for the training phase of RPC-Net. The experimental procedures adhered to the Declaration of Helsinki and were approved by the local ethics committee (CER-Polito, Prot. No. 107460/2023). Only DS1 was used in Section \ref{sec:variance}. In Section \ref{sec:RPC}, both data sets were used, but DS1 was modified by considering only 64 channels from the four proximal rows (instead of 96 channels). This was done to match the number of channels in DS2. DS1 was also used as data set in our previous work \cite{artgio}. Both datasets are available online \cite{zenodo_data}.

\subsubsection{Instrumentation}\label{sec:instr}

EMG was recorded on the surface of the dominant forearm using the MEACS system, an EMG amplifier developed at LISiN (Politecnico di Torino, Turin, Italy) \cite{Cerone2019}\cite{Cerone2021}. The system comprises multiple Sensor Units (SUs), each measuring 34 mm × 30 mm × 15 mm. The SUs sample 32 channels at 2.048 kHz (192 V/V gain, 16-bit resolution, 2.4 V dynamic range). The modular system can connect to various electrode arrays (e.g., HDE-Array), which, in this case, differed between DS1 and DS2. Hand position data were acquired using a VICON Motus motion capture system (VICON Motion Systems, Centennial, Oxford, United Kingdom) sampling at 100 samples per second. The setup included 14 infrared cameras (Vero v2.2). The MEACS system includes a wireless synchronization unit which connects to the VICON Lock. When motion tracking starts or stops, the unit emits a digital signal which is recorded alongside the EMG data for synchronization during post-processing.

\subsubsection{HDE-Array}\label{sec:array}

The HDE-Array is a 2D polyimide array assembled using elastic bands (Fig. \ref{fig:rpc_net_Design}). The unassembled array measures 632 mm by 48 mm, featuring 32 resin stiffeners spaced 20 mm apart. Two electrodes, 10 mm apart, are soldered on each stiffener, resulting in 64 electrodes arranged in 2 rows and 32 columns. The silver disc electrodes have a 4 mm diameter (12.57 mm$^{2}$ area). When assembled, the HDE-Array perimeter is 200 mm when not stretched.

\subsubsection{DS1: Subjects and Experimental Protocol}\label{sec:DS1}

Data were collected at the Laboratory of Motion Analysis (Politecnico di Torino, Turin, Italy) from 12 right-handed subjects (no surgical interventions on their dominant arm, forearm circumferences: 20–32 cm), seven males and five females (age: 20–26; weight: 55–90 kg; height: 165–195 cm). Written informed consent was obtained from all participants. Each session involved four trials where high-density EMG and hand position data were collected as subjects performed 16 hand poses: 4 wrist poses: flexion, extension, adduction, abduction; 8 finger poses: index finger metacarpal-phalangeal flexion and extension; index finger proximal interphalangeal flexion and extension; flexion and extension of the middle, ring, and little fingers; adduction and abduction of the index and middle fingers; 4 thumb poses: flexion, extension, adduction, abduction. Before each session, skin was scrubbed with an abrasive paste to lower skin-electrode impedance, and electrodes and markers were placed. Twenty-one reflective markers were positioned on the dominant hand (embedded in a glove), and 12 more on the upper limb and trunk, for a total of 33 markers (Fig. \ref{fig:mark_el_placement}). EMG was acquired using three SUs connected to anisotropic gel electrode arrays (2 rows, 16 columns, with 10 mm and 15 mm inter-electrode distances) for a total of N=96 monopolar EMG channels. Electrode placement followed \cite{artgio}, with electrodes arranged in 6 rows and 16 columns around the circumference of the forearm, covering approximately a third of its length. The reference electrode was placed on the lateral epicondyle. A 10-second recording was performed to calibrate the VICON system by instructing wrist movements. Each trial consisted of 48 movements (16 poses repeated three times). Participants stood with their dominant forearm on a vertical support at shoulder height. A monitor prompted poses in random order every 6–8 seconds. The interval between prompts was adjusted by the participants for comfort, without a specified transition speed. Each trial lasted 300-400 seconds, depending on the length of the interval between prompts.

\subsubsection{DS2: Subjects and Experimental Protocol}\label{sec:DS2}
\begin{figure}[t]
 \centering
 \includegraphics[width=\columnwidth]{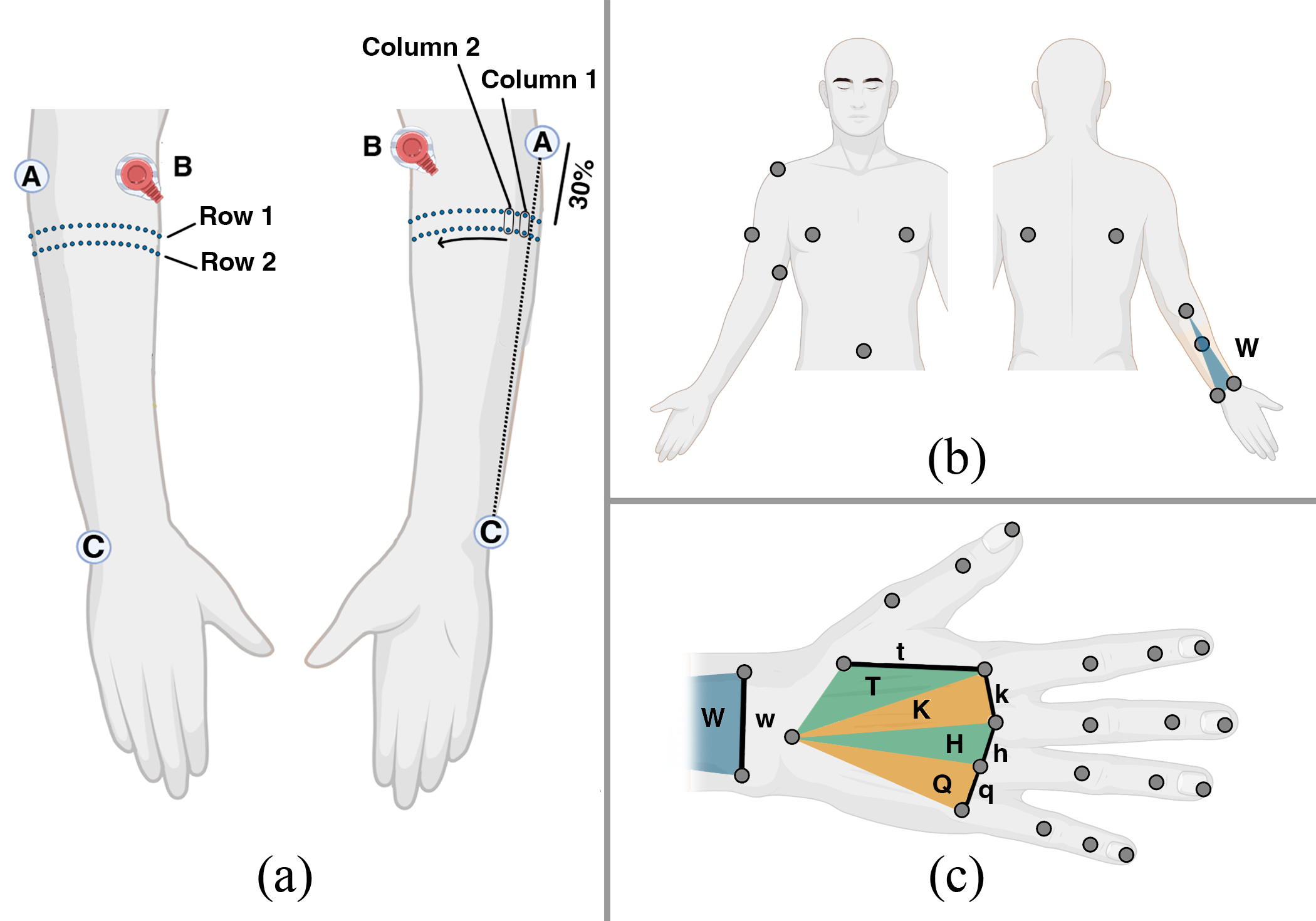}
 \caption{\textbf{Subject setup:} (a) HDE-Array placement on the front and back of the arm. Light blue dots indicate electrodes. 30\% of the distance is shown. (b) Placement of 12 infrared markers on the body of the subject. Plane W is highlighted in blue. (c) Placement of 21 infrared markers on the hand (and 2 forearm markers, also depicted in (b)). Markers are placed proximally to each joint. Planes W, K, T, H, and Q are highlighted in orange, green, and purple. Corresponding limiting lines are highlighted in black.}
 \label{fig:mark_el_placement}
\end{figure}

DS2 was collected at the Laboratory of Motion Analysis (Politecnico di Torino) between November and December 2023. Sixteen healthy volunteers (8 males, 8 females, age: 20–26; weight: 55–90 kg; height: 165–195 cm) participated. The protocol mirrored DS1, with some changes: 1) each movement was repeated twice, totaling 32 movements per trial (200–260 seconds per recording); 2) six trials were made per session; 3) two SUs were used with the HDE-Array, with dry electrodes arranged in 2 rows and 32 columns around the forearm (Fig. \ref{fig:mark_el_placement}). An additional elastic band around HDE-Array secured it and improved skin contact. The proximal row of electrodes was positioned at 30\% of the distance between the lateral epicondyle and pisiform bone.

\begin{figure}[t]
 \centering
 \includegraphics[width=\columnwidth]{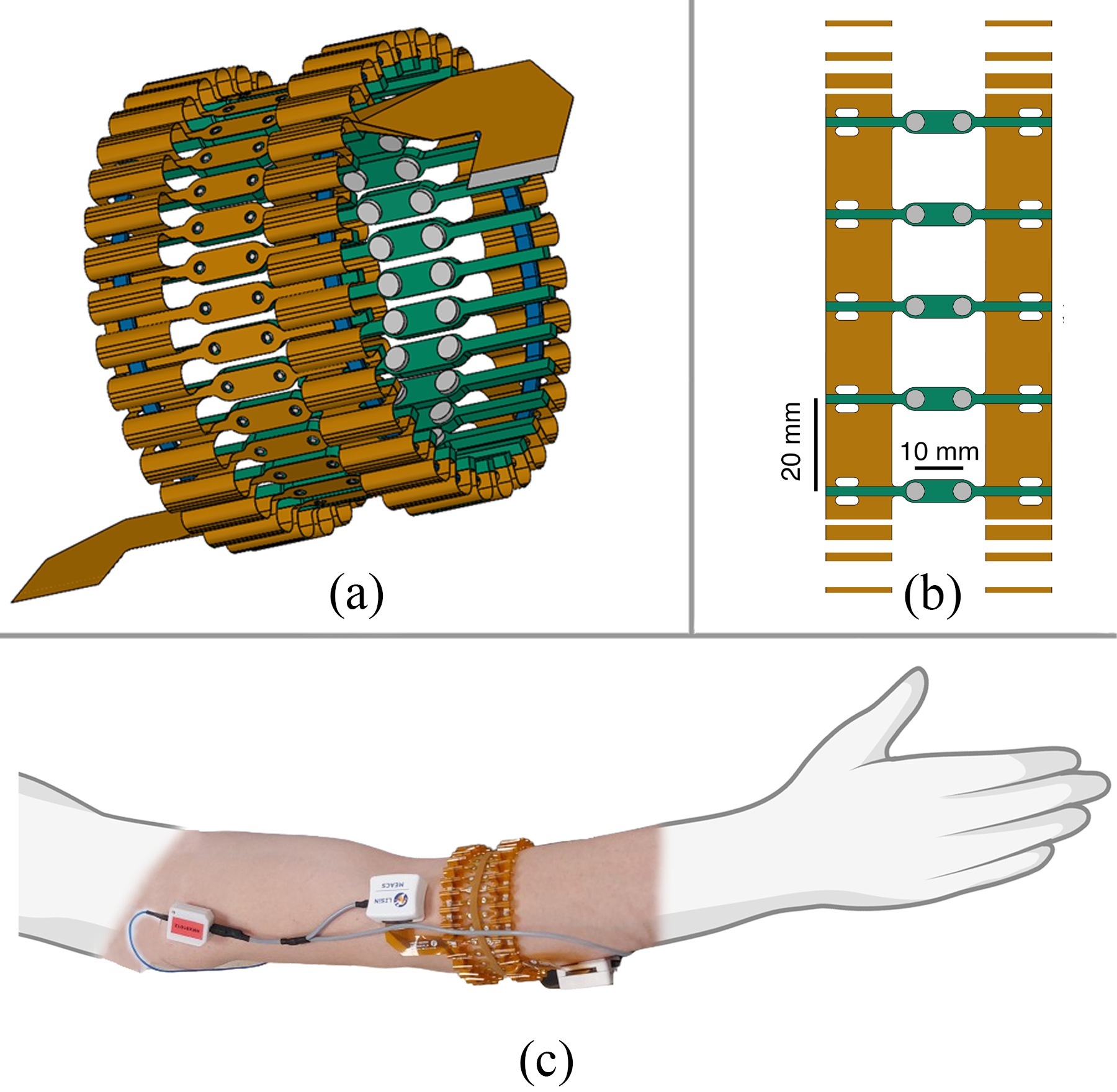}
 \caption{\textbf{HDE-Array:} (a) 3D rendering of the array in its assembled form. Polyimide is in orange, stiffeners in green, elastic bands in blue, and electrodes in silver. (b) Sketch of the repeating unit of the array (not assembled), including dimensions. (c) Electrode array positioning on the forearm of the subject, including EMG amplifier (in white).}
 \label{fig:rpc_net_Design}
\end{figure}

\subsection{Variance Analysis Across Columns and Rows for Gel Electrodes}\label{sec:variance}
The HDE-Array has increased circumferential electrode concentration, reducing the number of rows compared to typical configurations \cite{Chamberland2023}. To determine if this reduces information, we performed a variance analysis on DS1, following these steps for each subject: 1) Aggregate the four EMG recordings into a single 20-minute signal (n-by-96). 2) Compute the Root Mean Square (RMS) using a sliding 200-sample window to obtain signal B. 3) Normalize B by dividing by the channel-wise mean over time. 4) Reshape B into a n-by-16-by-6 signal C (16 rows and 6 columns). 5) Calculate variance across columns (proximo-distal) and rows (circumferential) for each time point of signal C, to get a n-by-16 signal D1 and a n-by-6 signal D2. 6) Average D1 and D2 over time to get a 1-by-16 array E1 and a 1-by-6 array E2. We define proximo-distal Normalized Dimensional Variance (NDV) as each of the values in E1 and circumferential NDV as each of the values in E2. This yields 264 data points: 192 proximo-distal and 72 circumferential. A Shapiro-Wilk test assessed distribution non-normality, followed by a one-tailed Mann-Whitney U test to evaluate the null hypothesis that median proximo-distal NDV is greater than or equal to median circumferential NDV. Rejection of this null hypothesis would indicate higher NDV across the circumferential direction \cite{witte2017statistics} \cite{casella2024statistical}. A higher NDV across the circumference would imply that a reduction in the number of rows does not result in a signal with significantly less information content, successfully addressing concerns of insufficient proximo-distal coverage.

\subsection{Skin-Electrode Impedance Characterization}\label{sec:impedance}

\begin{figure}[t]
 \centering
 \includegraphics[width=\columnwidth]{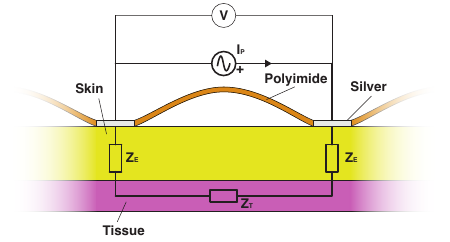}
 \caption{\textbf{Electrical model of the skin-electrode interface:} skin in yellow, electrode array in orange, electrodes in silver, and underlying tissue in pink. $Z_{T}$ is assumed to be much smaller than $Z_{E}$. Current $I_{P}$ is injected and voltage difference $V$ is measured.}
 \label{fig:impedance_circ}
\end{figure}

We characterized the electrode-skin impedance ($Z_{E}$) (Fig. \ref{fig:impedance_circ}) within the sEMG bandwidth. This is important, especially around power line frequency, because the interaction between the amplifier input stage and the electrode-skin system can cause signal distortion and attenuation due to the voltage divider effect and power line interference \cite{Merlettitutorial}\cite{Merletti2010elskin}\cite{Webster2020}\cite{Rijn1990}\cite{Merletti2010advance}\cite{Huhta1973}. Following a previous protocol, we measured impedance and noise using a multi-channel impedance meter from LISiN (Politecnico di Torino) \cite{Cerone2021}. For dry-contact silver electrodes, the electrode-skin contact can be modeled as a parallel R-C network. Within the sEMG bandwidth (10 Hz - 500 Hz), this network is predominantly non-polarizable and the impedance magnitude is primarily due to the resistive component and is less influenced by the reactive, frequency-dependent components \cite{Chi2010}\cite{Webster2020}\cite{Merletti2010elskin}. Measurements were taken between two identical electrodes, assuming comparable skin properties and negligible tissue impedance ($Z_{T}$) \cite{Piervirgili2014}. A schematic of the model circuit is shown in Fig. \ref{fig:impedance_circ}. To characterize $Z_{E}$, we conducted two analyses: 1) Impedance magnitude ($|Z_{E}|$) at 50 Hz was measured on 28 adjacent electrode pairs on the proximal ring of the HDE-Array for 5 subjects, totaling 140 measurements. For comparison, 28 measurements were also done on one subject using a gel electrode array (4-by-8, 10 mm inter-electrode distance, used for EMG acquisition in \cite{9668845}). Median and Inter-Quartile Range (IQR) were calculated. 2) For one subject, we measured impedance magnitude and phase across 50 frequencies (1 to 10,000 Hz, 12.3 points per decade) using 32 pairs of adjacent electrodes on the HDE-Array first and on a gel array (4-by-8, 10 mm spacing) then. A Bode plot was generated using the median phase and magnitude, with IQRs, for both conditions. These procedures aim to assess the reliability of the dry-electrode-skin interface for EMG acquisition, addressing concerns about high impedance of the HDE-Array. To further compare impedance differences, we recorded EMG during a bilateral isometric task. The subject lifted a 30 kg barbell for 12 seconds, while we recorded HD-sEMG from the right biceps (gel array, described above) and from the left biceps (HDE-Array). Spectrograms and RMS signals were computed for the channel on each arm corresponding to electrodes positioned on the mid-point between the acromion and biceps insertion \cite{barbero2012atlas}. We then calculated the error for the RMS signals (RMSE, mV) and for the spectrograms (RMSE, dB) to quantify differences between the two signals.

\subsection{RPC-Net Performance Assessment: HDE-Array and Gel Electrodes}\label{sec:RPC}
This experiment tested the main hypothesis by comparing RPC-Net performance using HD-sEMG signals from gel electrodes and the HDE-Array, verifying if the HDE-Array is suitable for position estimation. We used data from DS2 (HDE-Array) and DS1 (gel electrodes). Performance was defined as the accuracy of RPC-Net's position estimates compared to the ground truth, measured by the metrics in Section \ref{sec:measures}. EMG data were pre-processed, used to train RPC-Net, and then used by RPC-Net to return a position estimate whose accuracy is measured according to the metrics defined below. The process was conducted using MATLAB (R2023a, The MathWorks, Inc.) and VICON Nexus (v2.11, Oxford Metrics plc). Additional code was written in Python using BSD-licensed libraries.

\subsubsection{Pre-processing: HD-sEMG Data}\label{sec:emg_proc}
The EMG signal was pre-processed by converting from bits to volts, removing the offset, rectifying the signal, and computing RMS over a ($w_{l}$) 200-sample window (97.7 ms). The window slide was adjusted based on prompt duration, as defined in Section \ref{sec:DS1}: 25 samples for 8 seconds, 29 for 7 seconds, and 33 for 6 seconds, ensuring consistency in sample number across subjects. The RMS values were scaled by dividing by 10$^{-4}$ to normalise the data for RPC-Net training (null mean and unitary standard deviation). VICON position data were processed with a moving average filter (order 20) and mapped into a 29-dimensional joint-angle space using the Inverse Kinematic Algorithm (IKA) (Section \ref{sec:kin_mod}). The outputs were adjusted by subtracting rest angles and normalizing by 45 degrees to center the mean at zero and set standard deviation to one for network training. Linear interpolation aligned the position sampling rate with the processed EMG signals.

\subsubsection{Kinematic Model}\label{sec:kin_mod}
RPC-Net requires joint angles as input, so a kinematic model was developed to convert VICON marker positions to 29 hand joint angles. Based on models from \cite{Lee1995} and \cite{cobos2010human}, we developed a 29 DoF hand kinematic model. To translate marker positions to joint angles, we developed the IKA, based on the kinematic model. The core of the IKA is an optimization process designed to identify the 29 joint angles (one per kinematic DoF) that best approximate the marker positions, executed for each frame captured by the VICON system. The process has three phases: 1) wrist angles (3 DoF), 2) finger angles (5 DoF per finger), and 3) thumb angles (6 DoF). In phase 1, the three wrist angles represent hand rotations around the x, y, and z axes, starting from a reference position where the K plane is aligned with the W plane and lines k and w are parallel (Fig. \ref{fig:mark_el_placement}). In phase 2, each finger moves in flexion-extension at the metacarpal-phalangeal (MP) joint and at the interphalangeal (IP, ID) joints. The flexion plane for each finger is defined by two angles, $\alpha$ and $\beta$ (positive for pronation and radial deviation respectively), optimized using sequential quadratic programming (MATLAB; max evaluations=500; function tolerance=10$^{-1}$; optimality=10$^{-6}$) \cite{nocedal2006quadratic}. $\alpha$ and $\beta$ are zero when the flexion plane is perpendicular to the defining line ("k" for the middle and index finger, "h" for the ring, and "q" for the little). A sigmoid function such that $\alpha_{1}=\alpha_{0} * \text{sigmoid}(\zeta)$ and $\beta_{1}=\beta_{0} * \text{sigmoid}(\zeta)$ (where $\zeta$ represents the mean distance of the 4 markers from the principal direction of the same markers) refines $\alpha$ and $\beta$ to prevent excessive motion when the finger is not flexed. Markers are then projected onto the flexion plane, and three additional angles ($\gamma$, $\delta$, $\epsilon$), representing the angles between finger segments (MP, PIP, DIP), are computed. The three flexion angles are set to zero when the three finger markers lie on the K (index and middle), H (ring), or Q (little) planes. Flexion of a joint corresponds to a positive angle. In phase 3, the thumb moves in flexion-extension and adduction-abduction at the MP, PIP, and DIP joints. All thumb angles are set to zero when markers align on the T plane (perpendicular to the "t" line), and angles are computed by rotating the system to set the proximal joint at zero degrees. Phases 1 and 3 are error-free, while phase 2 introduces minor discrepancies between actual and estimated positions. Across all subjects, the average error in marker position was less than 1 mm (mean: 0.35 mm, standard deviation: 0.42 mm over 4,328,075 frames). The IKA projects from a 3D space to a J-dimensional joint angle space (J=29, number of joints). A Forward Kinematic Algorithm (FKA) was developed to convert joint angles into marker positions.

\subsubsection{RPC-Net Training and post-processing} \label{sec:trpppr}
RPC-Net is a neural network that estimates hand position from HD-sEMG signals acquired on the proximal forearm, refining estimates recursively based on prior values \cite{artgio}. The version of RPC-Net used in this study processes inputs from 64 EMG channels and 29 joint angles. Each sub-network, responsible for a specific joint angle, excludes its corresponding input channel, using the remaining 28 angles instead. This setup creates a robust recursive loop between EMG signals and past angles. During training, actual joint angles replace past estimates. Inference time (IT) is 5.4 ms (standard deviation 0.5 ms) over 10$^{5}$ iterations on an Intel Xeon Gold 5120 CPU @ 2.20GHz. RPC-Net was trained and tested independently for each subject with a 5:1 train-to-test ratio. The Adam optimizer (learning rate = 10$^{-5}$, $\varepsilon$ = 10$^{-3}$, $\beta_1$ = 0.9, $\beta_2$ = 0.99, batch size = 2000) and MSE Loss were used, both implemented in PyTorch. Training lasted 200 epochs. RPC-Net outputs were filtered using a low-pass Butterworth filter (order 6, cutoff frequency 1 Hz) and mapped into 3D marker positions using the FKA for comparison with the ground truth.

\subsubsection{Performance assessment} \label{sec:measures}

RPC-Net performance was evaluated for each subject using two metrics: Mean Pearson Correlation Coefficient (MPCC) and Mean Distance (MD). MPCC is the average Pearson Correlation Coefficient (PCC) comparing actual and predicted joint angles for the 29 DoFs during the test trial. Additionally, median, first, and third quartiles of PCCs were calculated. MD represents the average Weighted Fingertip Distance (WFD), defined as the average distance at each time point between the position of the tips of the index, middle and thumb finger and their position estimated by RPC-Net. The median, first and third quartiles of the WFD were computed over time. Performance is directly proportional to MPCC and inversely proportional to MD.

We also employed one-dimensional Statistical Parametric Mapping (1D SPM), designed to detect differences in pairs of time-varying signals \cite{pataky2012one}. Given the difference between the estimate of a signal and its value, the one-sample t-test computes a t-statistic at each time point, creating a test statistic map, which is then compared against a critical threshold derived from random field theory to assess statistical significance of the difference. The input is an n-by-k signal difference, where n is the number of time points in the time series, and k is the number of trials considered. The output is an n-by-1 t-statistic time series. We used the one-sample t-test (spm1d Python toolbox) to assess whether the distance between actual and estimated joint angles differed significantly from zero \cite{linkpataky}. Starting with 28 pairs (estimate and value) of test recordings (12 from DS1 and 16 from DS2), we computed the difference between actual and estimated 29 joint angles, resulting in 28 29-by-p recordings A, where p is the number of samples. These recordings were divided into 16 sub-recordings for the 16 movements, each lasting 8 seconds, resulting in 16 datasets B with 28 n-by-29 recordings each. The data in B were split into two groups, C1 and C2, for gel and dry electrodes respectively. We re-arranged C1 to obtain 29 n-by-12 signals D1 and C2 to obtain 29 n-by-16 signals D2, where 12 and 16 are the number of subjects in the gel and dry electrode datasets, respectively. We performed a one-sample t-test on each of the 29 signals D1, producing 29 t-statistic time series E1. The same was done for D2, yielding 29 t-statistic time series E2. Each t-test also returns a critical threshold value (significance=0.05). We subtracted each E1 and E2 from the corresponding critical threshold value to obtain as many F1 and F2. We then computed the average and interquartile range (IQR) of all F1 signals to get G1, an n-by-1 representation of cross-joint variability, termed Cross-Joint Derivative (CJD). The same was done for F2 signals to get G2, the CJD relative to the dry electrode dataset. This process produced two CJD signals (G1 and G2) for each dataset B (one per movement). Finally, we calculated the error difference between the gel and dry CJD signals for each movement, obtaining H (one value per movement). Averaging H across all movements returned a general measure of difference in performance between position estimates made with dry and gel electrodes, termed Cross-Movement, Cross-Joint measure of Difference (CMCJD).

\subsubsection{Comparative analyses} 

To test our hypothesis, we performed two comparative analyses: 1) We compared RPC-Net's performance (MD and MPCC) using signals from HDE-Array (32-by-2 electrodes) to its performance with gel electrodes (16-by-4 electrodes). 2) We assessed RPC-Net's performance with a subset of HDE-Array channels (16-by-2 configuration using alternate columns) versus the same subset from the gel array to evaluate dry and gel electrodes independently of their configuration (Fig. \ref{fig:all_res}). A one-tailed t-test was used to compare these conditions, testing the null hypothesis that RPC-Net's performance with the HDE-Array is inferior or equal to that with gel electrodes. MD and MPCC distributions were normal, as confirmed by a Shapiro-Wilk test. 

We also compared the performance of RPC-Net using EMG data from a single circumference of dry electrodes, testing individual rows and the full setup using three two-tailed t-tests. The null hypotheses were: 1) performance with both rows is equal to performance with the proximal row only, 2) performance with both rows is equal to performance with the distal row only, and 3) performance with the proximal row is equal to performance with the distal row. 

Lastly, we used the CJD to visualize the difference in accuracy between estimates performed with HD-sEMG acquired with gel and dry electrodes. For each of the 16 movements considered we plotted the CJD for the dry and gel group (along with the IQR) for a visual comparison. We also reported the CMCJD as an overall measure of difference.

\section{Results}

This section presents the findings from the variance analysis (Section \ref{sec:variance}), impedance validation (Section \ref{sec:impedance}), and RPC-Net performance (Section \ref{sec:RPC}).

\subsection{Variance Analysis}
\begin{figure}[t]
 \centering
 \includegraphics[width=\columnwidth]{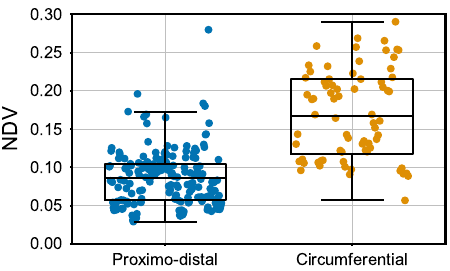}
 \caption{\textbf{Variance Analysis Results:} Distribution of proximo-distal and circumferential NDV, as outlined in Section \ref{sec:variance}. Mann-Whitney U test Results (H0: Median proximo-distal NDV $\geq$ Median circumferential NDV): $U$ = 1549, $n$ = 264, $p$ = 1.43e-22.}
 
 \label{fig:variance_results}
\end{figure}

Fig. \ref{fig:variance_results} shows the distribution of circumferential and proximo-distal NDV for EMG signals in DS1 without making a subject-wise distinction. Each data point represents an NDV measurement from individual sessions over time. The median proximo-distal NDV was 0.086 (IQR: 0.047), while the median circumferential NDV was 0.167 (IQR: 0.098). The Mann-Whitney U test results ($U$ = 1549, $n$ = 264, $p$ = 1.43e-22) indicate rejection of the null hypothesis.

\subsection{Impedance Characterization}
\begin{figure*}[t]
 \centering
 \includegraphics[width=\textwidth]{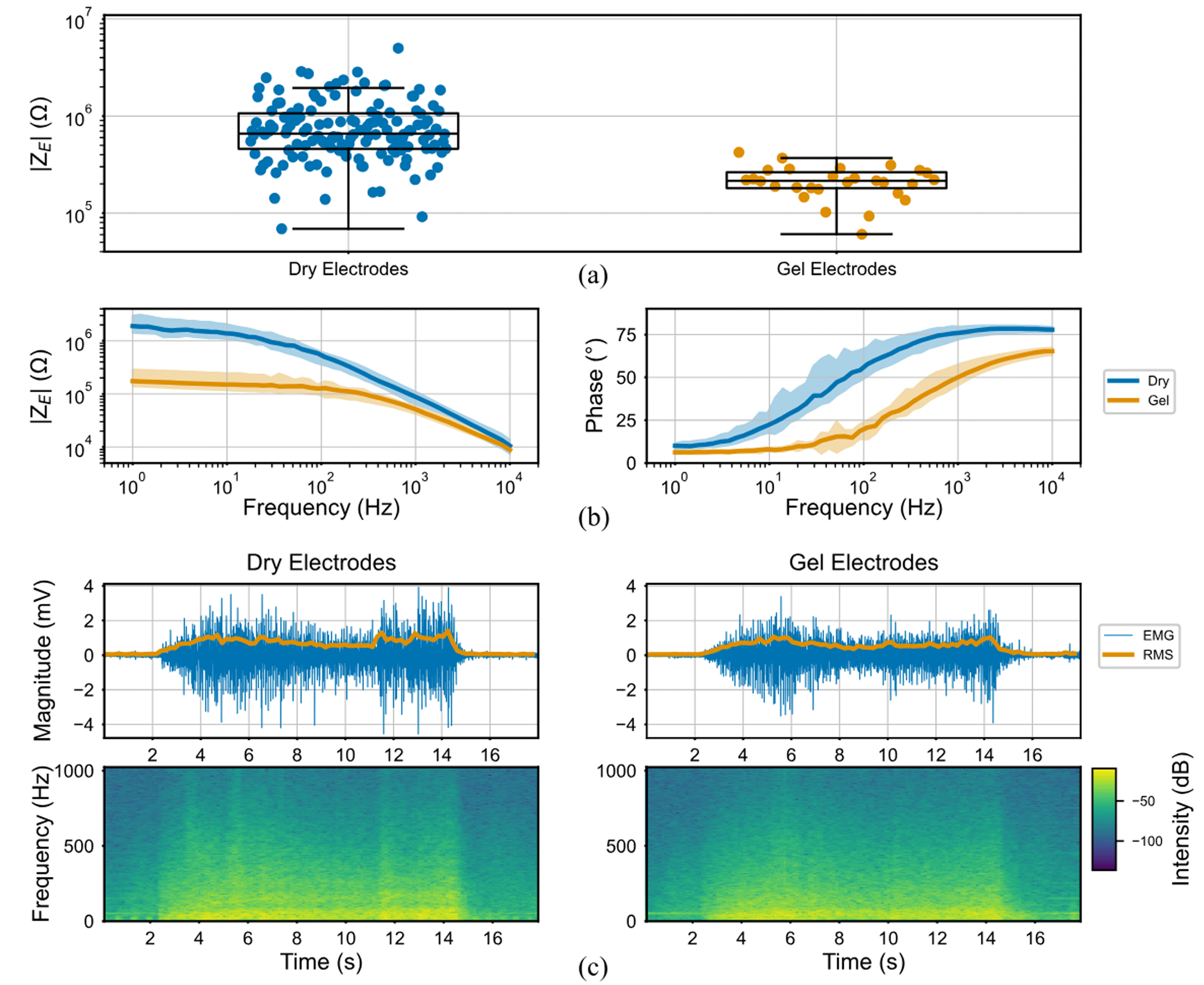}
 \caption{\textbf{Impedance Characterization Results:} The figure reflects the experimental conditions outlined in Section \ref{sec:impedance}. (a) Impedance values at 50 Hz. The y axis indicates $|Z_e|$, the magnitude of the impedance at 50 Hz (power line frequency). The x axis indicates the type of electrodes with which the measurement was taken. Measurements relating to the two conditions and the corresponding box plots are shown. Dry electrode median = 661 k$\Omega$ ($n$=140, First quartile: 460 k$\Omega$, Third quartile: 1064 k$\Omega$). Gel electrode median = 214 k$\Omega$ ($n$=28, First quartile: 180 k$\Omega$, Third quartile: 264 k$\Omega$). (b) Bode plot in the 1-10000 Hz range. For each frequency tested, the line represents the median across 32 measurements, and the shaded area the corresponding IQR. (c) EMG signals acquired from dry electrodes (left) and gel electrodes (right) during the same isometric task. The top row shows the raw EMG traces with the RMS values overlaid, while the bottom row presents the corresponding spectrograms for each condition. }
 \label{fig:electrical_validation}
\end{figure*}

Fig. \ref{fig:electrical_validation} shows impedance results (Section \ref{sec:impedance}). It displays the impedance values for both dry and gel electrodes at 50 Hz (Fig. \ref{fig:electrical_validation}.a). The median impedance amplitude observed at 50 Hz for dry electrodes is 661 k$\Omega$ (IQR: 604 k$\Omega$), which is almost three times higher than that for gel electrodes (214 k$\Omega$, IQR: 84 k$\Omega$). For a single subject, the Bode plot for the frequency range of 1-10000 Hz is shown (Fig. \ref{fig:electrical_validation}.b). Fig. \ref{fig:electrical_validation}.c shows comparable EMG signal frequency content and amplitude for both dry and gel electrodes. The raw signal traces in the top row show similar RMS profiles over the 12-second task, with overlapping RMS values, and an RMSE of 0.21 mV between RMS traces, confirming consistent signal power. The bottom-row spectrograms indicate similar frequency distributions for both electrode types, with an RMSE of 8.2 dB between the spectrograms.

\subsection{RPC-Net Performance} \label{sec:res_rpc}
\begin{figure*}[t]
 \centering
 \includegraphics[width=\textwidth]{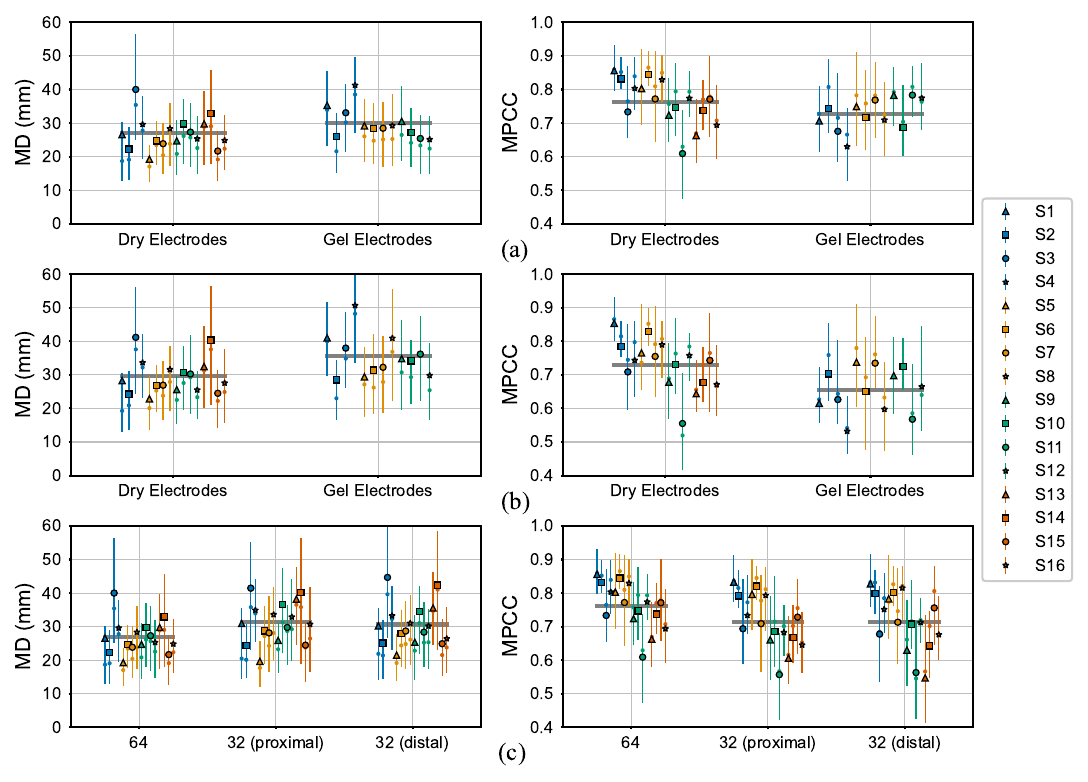}
 \caption{\textbf{RPC-Net Performance Results:} The figure reflects the experimental conditions outlined in Section \ref{sec:RPC}. (a) Performance indicators for RPC-Net across all subjects. One-sided paired t-test results for MD (H0: Gel Electrodes $\leq$ Dry Electrodes): $t(28)$ = - 1.63, $p$ = 5.8e-02. One-sided paired t-test results for MPCC (H0: Gel Electrodes $\geq$ Dry Electrodes): $t(28)$ = 1.51, $p$ = 7.2e-02. (b) Performance indicators for RPC-Net across all subjects, with a reduced number of electrodes for a more accurate comparison. One-sided paired t-test results for MD (H0: Gel Electrodes $\leq$ Dry Electrodes): $t(28)$ = - 2.73 $p$ = 5.6e-03. One-sided paired t-test results for MPCC (H0: Gel Electrodes $\geq$ Dry Electrodes): $t(28)$ = 2.78, $p$ = 4.9e-03. (c) Performance indicators for RPC-Net for dry electrodes only, comparing the full setup to one circumference at the time. One-sided paired t-test results for MD (H0: Full setup = Proximal Circumference): $t(16)$ = - 2.23, $p$ = 3.4e-02, (H0: Full setup = Distal Circumference): $t(16)$ = - 1.86, $p$ = 7.3e-02, (H0: Proximal Circumference = Distal Circumference): $t(16)$ = 0.3, $p$ = 7.65e-01. One-sided paired t-test results for MPCC (H0: Full setup = Proximal Circumference): $t(16)$ = 1.89, $p$ = 6.8e-02, (H0: Full setup = Distal Circumference): $t(16)$ = 1.79, $p$ = 8.3e-02, (H0: Proximal Circumference = Distal Circumference): $t(16)$ = 0.00, $p$ = 9.9e-01.}
 \label{fig:all_res}
\end{figure*}

This subsection compares the performance of the RPC-Net/HDE-Array setup with a traditional gel-based electrode setup and assesses the performance of single circumferences of HDE-Array electrodes (Section \ref{sec:RPC}). Fig. \ref{fig:all_res} depicts the subject-wise performance, measured as MD and MPCC. The caption includes statistical test results performed to assess the experimental hypotheses. Table \ref{table:all_res} presents the averages of MD and MPCC across subjects for each tested condition.

\begin{figure*}[t]
 \centering
 \includegraphics[width=\textwidth]{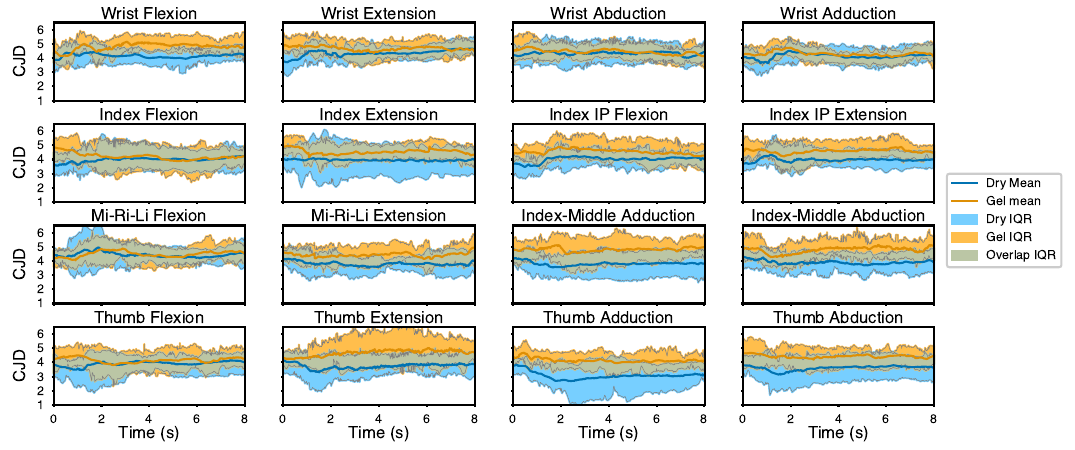}
 \caption{\textbf{SPM Analysis Results:} The figure reflects the conditions outlined in Section \ref{sec:RPC}. Comparison of CJD and corresponding IQR for the 16 movements considered, for position estimates obtained from data acquired with dry electrodes and gel electrodes. Critical t-value is computed for a significance of 0.05. A positive CJD indicates a less significant difference between value and estimate. }
 
 \label{fig:spmresults}
\end{figure*}

For the comparison between the 2-by-32 dry electrode setup and the 4-by-16 gel electrode setup, the mean MPCC was 0.76 for the dry electrodes and 0.73 for the gel electrodes, while the mean MD was 26.9 mm and 29.9 mm, respectively. The t-test results were as follows: MD: $t(28)$ = -1.63, $p$ = 5.8e-02; MPCC: $t(28)$ = 1.51, $p$ = 7.2e-02. For the simplified setup (2-by-16 electrodes), the mean MD was 29.5 mm for dry electrodes and 35.6 mm for gel electrodes ($t(28)$ = -2.73, $p$ = 5.6e-03), and the mean MPCC was 0.73 for dry electrodes and 0.65 for gel electrodes ($t(28)$ = 2.78, $p$ = 4.9e-03). The t-tests in both cases assessed the null hypothesis that the performance of RPC-Net using data acquired with gel electrodes is equal to or superior to the performance using data acquired with dry electrodes. Finally, when analyzing data from a single ring of electrodes, the t-tests showed differences between the full setup and the proximal ring setup. The mean MD was 26.9 mm for the full setup and 31.3 mm for the proximal ring ($t(16)$ = -2.23, $p$ = 3.4e-02), and the mean MPCC was 0.76 for the full setup and 0.71 for the proximal ring ($t(28)$ = 1.89, $p$ = 6.8e-02). Similarly, for the distal ring setup, the mean MD was 26.9 mm for the full setup and 30.6 mm for the distal ring ($t(16)$ = -1.86, $p$ = 7.3e-02), and the mean MPCC was 0.76 for the full setup and 0.71 for the distal ring ($t(28)$ = 1.79, $p$ = 8.3e-02). In this case, the t-test assessed the null hypothesis of equality of performance of RPC-Net when using data acquired with two rings and with a single ring. Fig. \ref{fig:spmresults} compares joint angle estimation accuracy across movements, showing the computed CJD for all movements. A higher CJD indicates a smaller difference between estimate and actual position, and zero is the threshold for statistical relevance. Gel electrodes showed higher accuracy, though values are consistently greater than zero for both signals. CMCJD is 0.82.

\begin{table}[h]
\caption{RPC-Net Performance}
\setlength{\tabcolsep}{3pt}
\renewcommand{\arraystretch}{1.2}

\begin{tabular}{|c|c|c|}

\hline

\rule{0pt}{10pt}\raisebox{1pt}{Condition} & \raisebox{1pt}{MD} & \raisebox{1pt}{MPCC}\\
\hline
Dry - 32-by-2 & 26.9 (mm) & 0.76 \\
\hline
Gel - 16-by-4 & 29.9 (mm) & 0.73 \\
\hline
Dry - 16-by-2 & 29.5 (mm) & 0.73\\
\hline
Gel - 16-by-2 & 35.6 (mm) & 0.65 \\
\hline
Dry - 32-by-1 (proximal) & 31.3 (mm) & 0.71 \\
\hline
Dry - 32-by-1 (distal) & 30.6 (mm) & 0.71 \\
\hline

\multicolumn{3}{p{246pt}}{Performance of RPC-Net. The mean of MD and MPCC (across subjects) is shown.}

\end{tabular}
\label{table:all_res}
\end{table}

\section{Discussion}

HDE-Array is a viable device for acquiring HD-sEMG data for accurate hand position estimation through RPC-Net.

In the first part of the study, variance analysis of EMG signals from the 16-by-6 setup showed greater variability around the forearm circumference than along the proximo-distal axis, confirmed by the low p-value from the Mann-Whitney U test. This result aligns with the longitudinal structure of forearm muscles \cite{drake2012gray}. Reducing rows to increase circumferential electrodes doesn't significantly compromise information. The HDE-Array achieves double the circumferential resolution of similar devices and four times that of the Myo armband \cite{chamberland2023novel} \cite{myojs}. HDE-Array also allows for increased proximo-distal channels by adding another array (32-by-4 electrodes).

Secondly, we characterized the skin-electrode impedance of the dry electrodes. At 50 Hz, the impedance (661 k$\Omega$, IQR: 604 k$\Omega$) was higher than that of gel electrodes (214 k$\Omega$, IQR: 84 k$\Omega$) but still well below the amplifier's input impedance at the same frequency (80 M$\Omega$) \cite{Cerone2019}. The normalized impedance (median 83 k$\Omega$ cm²) aligns with the theoretical estimate of 58.50 ± 64.16 k$\Omega$ cm² from \cite{li2017towards}. The Bode plots (Fig. \ref{fig:electrical_validation}.b) confirm that the skin-electrode impedance can be modeled as a parallel R-C network, with no significant non-linearities at significant EMG frequencies (10-500 Hz) for either dry or gel electrodes. The higher IQR for dry electrodes suggests variability due to differences in contact pressure or other factors, indicating variable impedance at the skin-electrode interface. The spectrograms and raw signal traces (Fig. \ref{fig:electrical_validation}.c) show that the quality of EMG signals remains comparable to gel electrodes, as demonstrated by the low overall errors between traces and between spectrograms. Overall the dry electrodes, although exhibiting higher impedance across the frequency spectrum, have electrical properties suitable for HD-sEMG acquisition, consistently with similar studies \cite{Cerone2021} \cite{cattarello2016characterization}.

Results in Fig. \ref{fig:all_res}.a and Section \ref{sec:res_rpc} show that RPC-Net with dry electrodes performs at least as well, if not better, than with gel electrodes in terms of MPCC and MD. One-tailed t-tests give p-values between 0.1 and 0.05, suggesting a weak rejection of the null hypothesis, with a slight trend favoring dry electrodes. However, under a two-tailed test, the p-values double to 0.116 and 0.144, meaning the hypothesis of equal performance cannot be rejected. Fig. \ref{fig:all_res}.b further supports this, showing RPC-Net performs better with dry electrodes even with the same array disposition (2-by-16), with p-values below 0.05. These results demonstrate that the performance of RPC- Net remains higher with dry electrodes not just due to increased circumferential resolution. Results from single circumferences (proximal or distal) are slightly worse but comparable to the full setup, as indicated by t-tests with p-values between 0.03 and 0.09. Although there’s a weak rejection of equal performance, MPCC and MD remain within acceptable limits for single-ring setups. This result is important as it allows for a 75\% decrease in time (1.2 ms) and a simpler, cheaper setup using a single MEACS SU. Fig. \ref{fig:spmresults} and Section \ref{sec:res_rpc} highlight that the performance of RPC-Net with dry electrodes is comparable to that with gel electrodes. In both cases, and consistently across the 16 movements considered, the CJD is greater than zero, indicating that the difference between actual and estimated positions is not statistically significant. The low CMCJD (0.82) further confirms the similarity in behavior between the two setups.

 All these results demonstrate our experimental hypothesis: it is possible to replicate the results observed with RPC-Net and gel electrodes using the newly designed HDE-Array. With similar performance, the added value of the RPC-Net/HDE-Array System is given by the enhanced comfort, acceptance and usability without compromising control accuracy. The full HDE-Array and RPC-Net setup outperform, or perform as well as, all other solutions in the academic literature in terms of offline performance, including those returned by RPC-Net when used with gel electrodes \cite{simpetru2023proportional}\cite{simpetru2022accurate}\cite{artgio}\cite{chamberland2023novel}.

\section{Conclusion}
This study introduced the novel HDE-Array for HD-sEMG acquisition using dry electrodes. We demonstrated that this setup performs comparably to traditional gel-based electrode setups when data acquired with it are used for position estimation through RPC-Net. We also assessed the electrical properties of dry electrodes and found them suitable for EMG acquisition. These findings suggest that HDE-Array, combined with RPC-Net, offers a viable alternative to traditional gel-based setups, potentially improving the practicality and comfort of HD-sEMG acquisition for prosthetic control and other applications in rehabilitation technologies. The improved usability brought by dry electrodes can make the technology more accessible for daily use, reducing the preparation time and discomfort associated with gel electrodes. Future work will focus on further refining HDE-Array design and conducting real-world testing with prosthetic devices and FES. Additionally, we aim to explore applications of this system beyond the medical field, recognizing that such a development could have numerous applications in the broader domain of human-computer interaction. In conclusion, HDE-Array and RPC-Net combination represents a significant advancement in the field of HD-sEMG acquisition, promising an enhanced user experience and broad applicability in biomedical engineering.

\section{Acknowledgment}

The authors would like to thank the members of ONIG in Oxford and of LISiN in Turin for the valuable guidance provided during the writing of this article and to acknowledge the use of the University of Oxford Advanced Research Computing (ARC) facility in carrying out this work \cite{richards2015university}. Fig. \ref{fig:mark_el_placement} and Fig. \ref{fig:rpc_net_Design} were created with BioRender.com.


\end{document}